\documentclass[aps,prl,twocolumn,reprint,superscriptaddress]{revtex4-1}
\usepackage{graphicx}
\usepackage{amsmath}
\usepackage{upgreek}
\usepackage{setspace}
\usepackage{braket}
\begin{document}

\title{Transform-limited photons from a coherent tin-vacancy spin in diamond}

\author{Matthew E. Trusheim*}
\affiliation{Department of Electrical Engineering and Computer Science, Massachusetts Institute of Technology, Cambridge, Massachusetts 02139, USA}
\author{Benjamin Pingault*}
\affiliation{Cavendish Laboratory, University of Cambridge, JJ Thomson Avenue, Cambridge CB3 0HE, United Kingdom}
\author{Noel H Wan}
\affiliation{Department of Electrical Engineering and Computer Science, Massachusetts Institute of Technology, Cambridge, Massachusetts 02139, USA}
\author{Mustafa G\"{u}ndo\u{g}an}
\email[]{Current address: Institut f\"{u}r Physik, Humboldt-Universit\"{a}t zu Berlin, 12489 Berlin, Germany}
\affiliation{Cavendish Laboratory, University of Cambridge, JJ Thomson Avenue, Cambridge CB3 0HE, United Kingdom}
\author{Lorenzo De Santis}
\affiliation{Department of Electrical Engineering and Computer Science, Massachusetts Institute of Technology, Cambridge, Massachusetts 02139, USA}
\author{Romain Debroux}
\affiliation{Cavendish Laboratory, University of Cambridge, JJ Thomson Avenue, Cambridge CB3 0HE, United Kingdom}
\author{Dorian Gangloff}
\affiliation{Cavendish Laboratory, University of Cambridge, JJ Thomson Avenue, Cambridge CB3 0HE, United Kingdom}
\author{Carola Purser}
\affiliation{Cavendish Laboratory, University of Cambridge, JJ Thomson Avenue, Cambridge CB3 0HE, United Kingdom}
\author{Kevin C. Chen}
\affiliation{Department of Electrical Engineering and Computer Science, Massachusetts Institute of Technology, Cambridge, Massachusetts 02139, USA}
\author{Michael Walsh} 
\affiliation{Department of Electrical Engineering and Computer Science, Massachusetts Institute of Technology, Cambridge, Massachusetts 02139, USA}
\author{Joshua J. Rose}
\affiliation{Cavendish Laboratory, University of Cambridge, JJ Thomson Avenue, Cambridge CB3 0HE, United Kingdom}
\author{Jonas N. Becker} 
\affiliation{Clarendon Laboratory, University of Oxford, Parks road, Oxford, OX1 3PU, United Kingdom}
\author{Benjamin Lienhard}
\affiliation{Department of Electrical Engineering and Computer Science, Massachusetts Institute of Technology, Cambridge, Massachusetts 02139, USA}
\author{Eric Bersin}
\affiliation{Department of Electrical Engineering and Computer Science, Massachusetts Institute of Technology, Cambridge, Massachusetts 02139, USA}
\author{Ioannis Paradeisanos}
\affiliation{Cambridge Graphene Centre, University of Cambridge, Cambridge CB3 0FA, UK}
\author{Gang Wang}
\affiliation{Cambridge Graphene Centre, University of Cambridge, Cambridge CB3 0FA, UK}
\author{Dominika Lyzwa}
\affiliation{Department of Electrical Engineering and Computer Science, Massachusetts Institute of Technology, Cambridge, Massachusetts 02139, USA}
\author{Alejandro R-P. Montblanch}
\affiliation{Cavendish Laboratory, University of Cambridge, JJ Thomson Avenue, Cambridge CB3 0HE, United Kingdom}
\author{Girish Malladi}
\affiliation{College of Nanoscale Science and Engineering, Suny Poly, 257 Fuller Road, Albany, NY 12203, United States}
\author{Hassaram Bakhru}
\affiliation{College of Nanoscale Science and Engineering, Suny Poly, 257 Fuller Road, Albany, NY 12203, United States}
\author{Andrea C. Ferrari}
\affiliation{Cambridge Graphene Centre, University of Cambridge, Cambridge CB3 0FA, UK}
\author{Ian A. Walmsley}
\affiliation{Clarendon Laboratory, University of Oxford, Parks road, Oxford, OX1 3PU, United Kingdom}
\author{Mete Atat\"{u}re}
\email[]{ma424@cam.ac.uk}
\affiliation{Cavendish Laboratory, University of Cambridge, JJ Thomson Avenue, Cambridge CB3 0HE, United Kingdom}
\author{Dirk Englund}
\email[]{englund@mit.edu}
\affiliation{Department of Electrical Engineering and Computer Science, Massachusetts Institute of Technology, Cambridge, Massachusetts 02139, USA}

\date{\today}
\begin{abstract}
Solid-state quantum emitters that couple coherent optical transitions to long-lived spin qubits are essential for quantum networks. Here we report on the spin and optical properties of individual tin-vacancy (SnV) centers in diamond nanostructures. Through cryogenic magneto-optical and spin spectroscopy, we verify the inversion-symmetric electronic structure of the SnV, identify spin-conserving and spin-flipping transitions, characterize transition linewidths, measure electron spin lifetimes and evaluate the spin dephasing time. We find that the optical transitions are consistent with the radiative lifetime limit even in nanofabricated structures. The spin lifetime is phonon-limited with an exponential temperature scaling leading to $T_1$ $>$ 10 ms, and the coherence time, $T_2^*$ reaches the nuclear spin-bath limit upon cooling to 2.9 K. These spin properties exceed those of other inversion-symmetric color centers for which similar values require millikelvin temperatures. With a combination of coherent optical transitions and long spin coherence without dilution refrigeration, the SnV is a promising candidate for feasable and scalable quantum networking applications.

\end{abstract}

\pacs{}

\maketitle
A central goal of quantum information processing is the development of quantum networks consisting of stationary, long-lived matter qubits coupled to flying photonic qubits~\cite{Kimble2008-cu, Wehner2018-wx}, with applications in quantum computing, provably secure cryptography, and quantum-enhanced metrology~\cite{Awschalom2018-en}. Among matter qubits, quantum emitters in wide-bandgap semiconductors~\cite{Gao2015-rj,Atature2018-oh} have emerged as leading systems as their coherent, spin-selective optical transitions act as an interface between quantum information stored in their spin degrees of freedom and emitted photons. While most work has so far focused on the nitrogen-vacancy (NV) center in diamond~\cite{Togan2010-fn,Childress2013-yy,Kalb2017-rf}, its low percentage of emission into the coherent zero-phonon-line (ZPL)~\cite{Riedel2017-ft} and large spectral diffusion when located near surfaces~\cite{Faraon2012-gz, Siyushev2013-kv}, fueled the investigation of alternative emitters. These include the group-IV color centers in diamond~\cite{Thiering2018-rt}, comprising the silicon-vacancy (SiV)~\cite{Hepp2014-yl,Rogers2014-bx,Sipahigil2016-xm,Rose2018-iq}, germanium-vacancy (GeV)~\cite{Siyushev2017-lp,Bhaskar2017-uq}, and the recently observed lead-vacancy (PbV)~\cite{Trusheim2018-jm} centers. These centers have a large fraction of emission into the ZPL and a crystallographic inversion symmetry that limits spectral diffusion and inhomogeneous broadening~\cite{Rogers2014-vl,Evans2016-ay}. Unlike the NV center, however, the electronic spin coherence of SiV and GeV centers is limited by phonon scattering to an upper-lying ground-state orbital~\cite{Jahnke2015-pu,Pingault2017-gs}, requiring operation at dilution-refrigerator temperatures ($<$ 100 mK)~\cite{Sukachev2017-sv,Becker2018-nv}, or controllably induced strain~\cite{Sohn2018-bb} to achieve long coherence times. 

The tin-vacancy (SnV) center in diamond~\cite{Iwasaki2017-mg,Tchernij2017-ba} is a group-IV color center that promises favorable optical properties and long spin coherence time at significantly higher temperatures ($>1$ K). Electronic structure calculations predict that the SnV has the same symmetry as the SiV and GeV\cite{Thiering2018-rt,Riedel2017-ft}, while experimental measurement of a large ground-state orbital splitting indicates that single-phonon scattering, the dominant spin dephasing mechanism of SiV and GeV centers at liquid helium temperatures, should be suppressed significantly~\cite{Iwasaki2017-mg}. Here we report spectroscopic measurements that are consistent with the conjectured electronic structure of the SnV, demonstrate that its optical transitions are radiative lifetime-limited, show optical initialization and readout of its electronic spin, and, finally, demonstrate improved spin decay times compared to SiV and GeV centers in similar conditions. These single-emitter measurements indicate the strong potential of the SnV for quantum optics and quantum networking applications. 

We first perform cryogenic magneto-optical spectroscopy on single SnV centers created through Sn implantation in an ultra-pure diamond grown by chemical vapor deposition~\cite{SnVSuppInfo} to determine the electronic properties of the SnV. The SnV energy spectrum is analogous to those of other group IV centro-symmetric color centers, such as the SiV and GeV centers. Its center of symmetry, located at the impurity atom (Sn, Ge, Si), leads to the color center having no permanent electric dipole. This makes it insensitive to first-order fluctuations in electric field and thus favours spectral stability and emission into the zero-phonon line. This symmetry results in doubly degenerate ground and excited orbital states, as depicted in Fig. 1a, which are split by spin-orbit interaction leading to characteristic optical transitions consisting of two doublets. At 4 K, the upper branch of the excited state rapidly thermalizes to the lower branch, leading to only the transitions labeled $\gamma$ and $\delta$ being visible under non-resonant excitation, as shown in Fig. 1b. Figure 1c displays the evolution of the optical spectrum of a single emitter when a magnetic field is applied along the [001] crystallographic direction of the diamond (corresponding to an angle of $54.7^{\circ}$ with respect to the high symmetry axis of the SnV along the $<$111$>$ directions). The visible transitions $\gamma$ and $\delta$ each split into four lines, analogously to the SiV and GeV centers, as expected for an electronic spin-1/2 system.

\begin{figure*}
    \includegraphics[width=\textwidth]{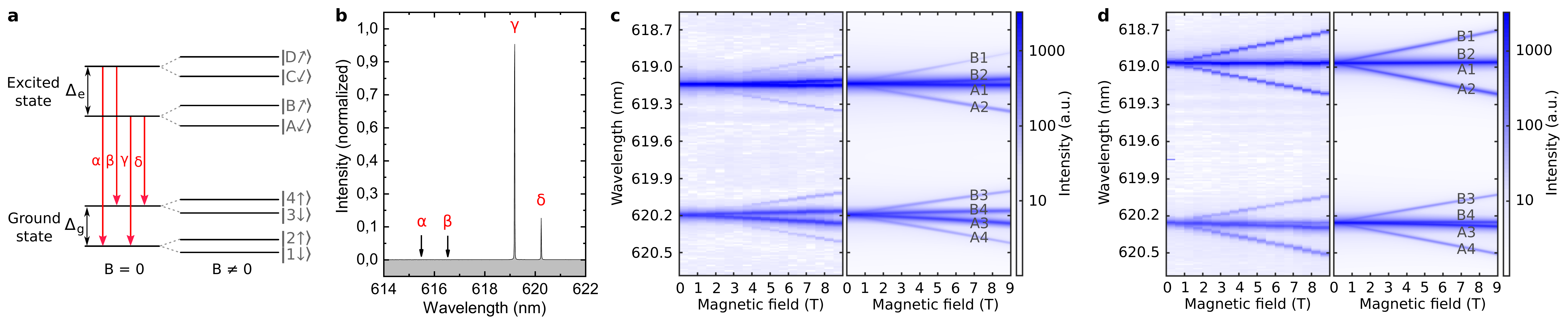}
    \caption{SnV center electronic structure. (a) The SnV energy levels, in the absence of magnetic field, are composed of doubly-split ground and excited states with respective splittings $\Delta_g$ and $\Delta_e$, resulting in four optical transitions (red arrows) labeled from $\alpha$ to $\delta$. A magnetic field splits each level into two. The spin state expected from the SnV model (see main text) is indicated for each level with up and down gray arrows. The tilted arrows in the excited state represent the difference in spin quantization axis compared to the ground state. (b) Fluorescence spectrum of a single SnV center at 4 K without magnetic field. (c) Left panel: Magnetic field-dependent fluorescence spectrum of a single unstrained SnV center. Right panel: simulated magnetic field dependence based on the model developed for the SiV center \cite{Hepp2014-yl}. The transitions are labeled according to the participating states. The simulated transition intensities were obtained from Fermi's golden rule for dipolar electric transitions and thus depend on the orbital and spin components of the initial and final states. (d) Magnetic field dependence of the fluorescence spectrum of a strained SnV center, both experimental (left panel) and simulated using the strain simulation developed for SiV in previous work~\cite{heppThesis, Meesala2018-zl} (right panel).}
    \label{fig1}
\end{figure*}
Owing to the common symmetry of group-IV color centers, the optical emission spectra can be simulated using the same group-theoretical model developed for SiV~\cite{Hepp2014-yl}. This model includes the Jahn-Teller effect and inherent pre-strain (such effects have similar Hamiltonians), spin-orbit coupling, and the orbital and spin Zeeman effects. Fitting the model to the experimental spectra, yields the simulated spectra displayed on the right panel of Fig. 1c. From this simulation, we determine that the $\sim$850 GHz ground-state splitting for unstrained emitters in the absence of magnetic field is almost entirely due to spin-orbit coupling (99\%). Spin-orbit interaction accounts for only 80\% of the $\sim$3000 GHz excited state splitting, with the rest being due to the Jahn-Teller effect. These couplings impact the electronic spin state associated with energy levels: the spin has an effective quantization axis defined by the combination of the orientation of the external magnetic field and the spin-orbit coupling, which tends to align the spin along the SnV high symmetry axis. As a consequence, in the range of magnetic field strength applied ($<$ 9 T), the spin quantization axis is dominated by the spin-orbit coupling. However, since the relative strengths of the spin-orbit coupling, Jahn-Teller effect, and magnetic field differ between the ground and excited states, the orientation of the effective quantization axes in the ground and excited states differs slightly. This leads, as shown in Fig. 1c, to dominant optical transitions between states with largely overlapping spin orientations (B2, A1, B4, A3), and weak transitions between states of nearly opposite spin orientations (B1, A2, B3, A4). 

Strain, likely induced during implantation, affects the spectrum of the SnV similarly to the SiV: it alters the orbital components of the states and thus the spin quantization axes via spin-orbit coupling. In our sample we observe many SnVs under large strain, indicated by an inhomogeneous distribution with a width on the order of THz~\cite{SnVSuppInfo}. As a consequence, the optical selection rules are relaxed and emission into the weaker transitions is reinforced. We observe this trend in strained SnV spectra (Fig. 1d), where the observed intensity of the weaker transitions is an order of magnitude greater than in the unstrained case (Fig. 1c). The spectra for the strained sample is consistent with a numerical model that adds a strain contribution to the emitter Hamiltonian~\cite{heppThesis, Meesala2018-zl,SnVSuppInfo}. These findings elucidate the photophysics of the SnV center and confirm its similarity with the previously described group-IV centers.

Following the determination of the SnV electronic structure, we investigate the coherent optical properties of individual SnV centers in nanofabricated pillars. Figure 2a shows a confocal scan of SnV centers in pillars with radius R = 150 nm under non-resonant 532 nm excitation. After identifying a SnV center through non-resonant scans, we perform photoluminescence excitation spectroscopy (PLE) by tuning a narrowband laser across the SnV $\gamma$ optical transition and detecting fluorescence in the phonon sideband. Across several single SnV centers, we measure a mean linewidth (Lorentzian FWHM) of $57 \pm 17$ MHz~\cite{SnVSuppInfo}, with some emitters displaying linewidths down to $30 \pm 2$ MHz (Fig. 2b). As a further demonstration of optical coherence, we measure the second-order photon correlation function g$^{(2)}$ for the emitter presented in Fig. 2b under resonant excitation and an additional continuous low-power non-resonant repump at 532 nm (Fig. 2c)~\cite{SnVSuppInfo}. The characteristic antibunching g$^{(2)}$(0) = $0.09~\pm~0.03$ confirms single photon emission under resonant excitation. At a resonant optical power above saturation, we furthermore observe Rabi oscillations in the g$^{(2)}$ with a decay time of $5.9 \pm 0.3$ ns, indicating coherence between the driven levels on this timescale. Finally, we measure the fluorescence lifetime of this SnV center after pulsed non-resonant excitation at 532 nm (Fig. 2d). We observe an exponential decay of fluorescence with a time constant $\tau$ = $4.5\pm 0.2$ ns. Together with the transition linewidth measurement, this indicates that SnV optical transitions are consistent with the lifetime limit of $(2\pi\tau)^{-1} = 35 \pm 5$ MHz.  The mean transition linewidth is within a factor of 1.5 of the transform limit. Lifetime-limited optical transitions demonstrate the high optical quality of the SnV and are essential for its use as an optically interfaced spin qubit.

\begin{figure}
    \includegraphics[width = 8.5cm]{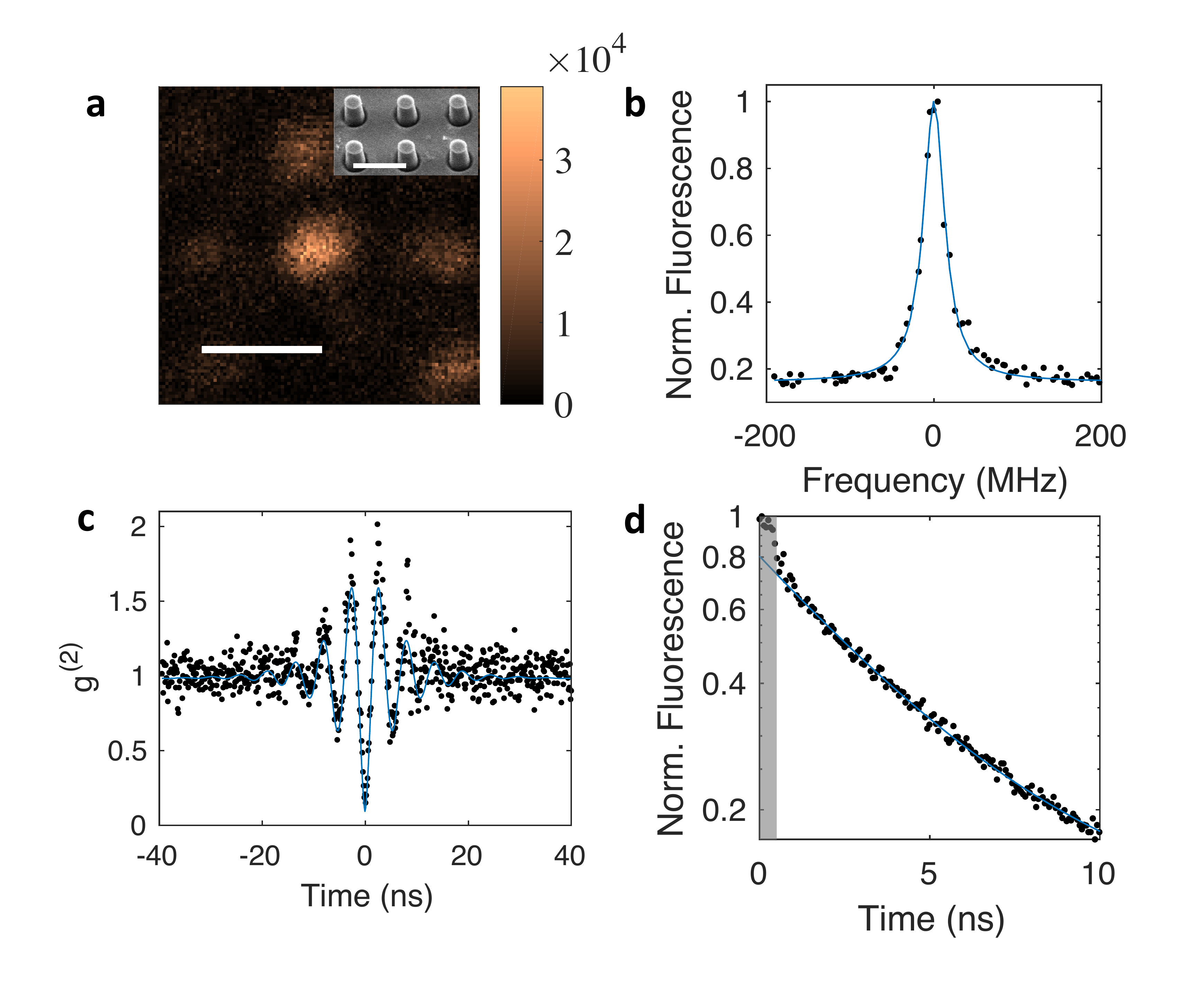}
    \caption{SnV center optical coherence. (a) Confocal microscope image of SnV centers in nanopillars with radius r = 150 nm (inset, scanning electron micrograph). Scale bars are 1 $\upmu$m. (b) Resonant PLE spectrum of a single SnV optical transition. Each point was integrated for 0.6 seconds. Blue line: Lorentzian fit with FHWM of $30 \pm 2$ MHz. (c) Second-order correlation function under resonant driving. (d) Fluorescence decay after non-resonant pulsed excitation at $t=0$. The gray shaded area between 0 and 0.5 ns indicates the instrument response.  
}
    \label{fig2}
\end{figure}

To probe the expected spin degree of freedom of the unstrained SnV, we implement spin-selective resonant excitation in an external magnetic field. Driving the transition A1 under a 9~T magnetic field oriented along the [001] crystallographic direction populates the Zeeman sublevel A, as depicted in Fig. 3a, resulting in emission predominately through transition A3 (Fig. 3b). Conversely, driving transition B2 pumps population into level B and results in emission through B4. This dependence shows evidence of selection rules in the thermalization mechanism within the excited state~\cite{Muller2014-ba} confirming that the SnV possesses a spin 1/2 and demonstrating its optical accessibility. As a second step, we tune the laser frequency into resonance with the weak transitions B1 and A2 between levels of differing spin states, as depicted in Fig. 3c. The spin selection rules would prohibit excitation, however, the observation of emission through transitions B4 and A3 confirms relaxation of the optical selection rules due to the different effective spin quantization axes between ground and excited states. This observation of spin selection rules also confirms the assignment of the experimentally measured transitions to the energy levels of the model.

\begin{figure}
    \includegraphics[width = 8.5cm]{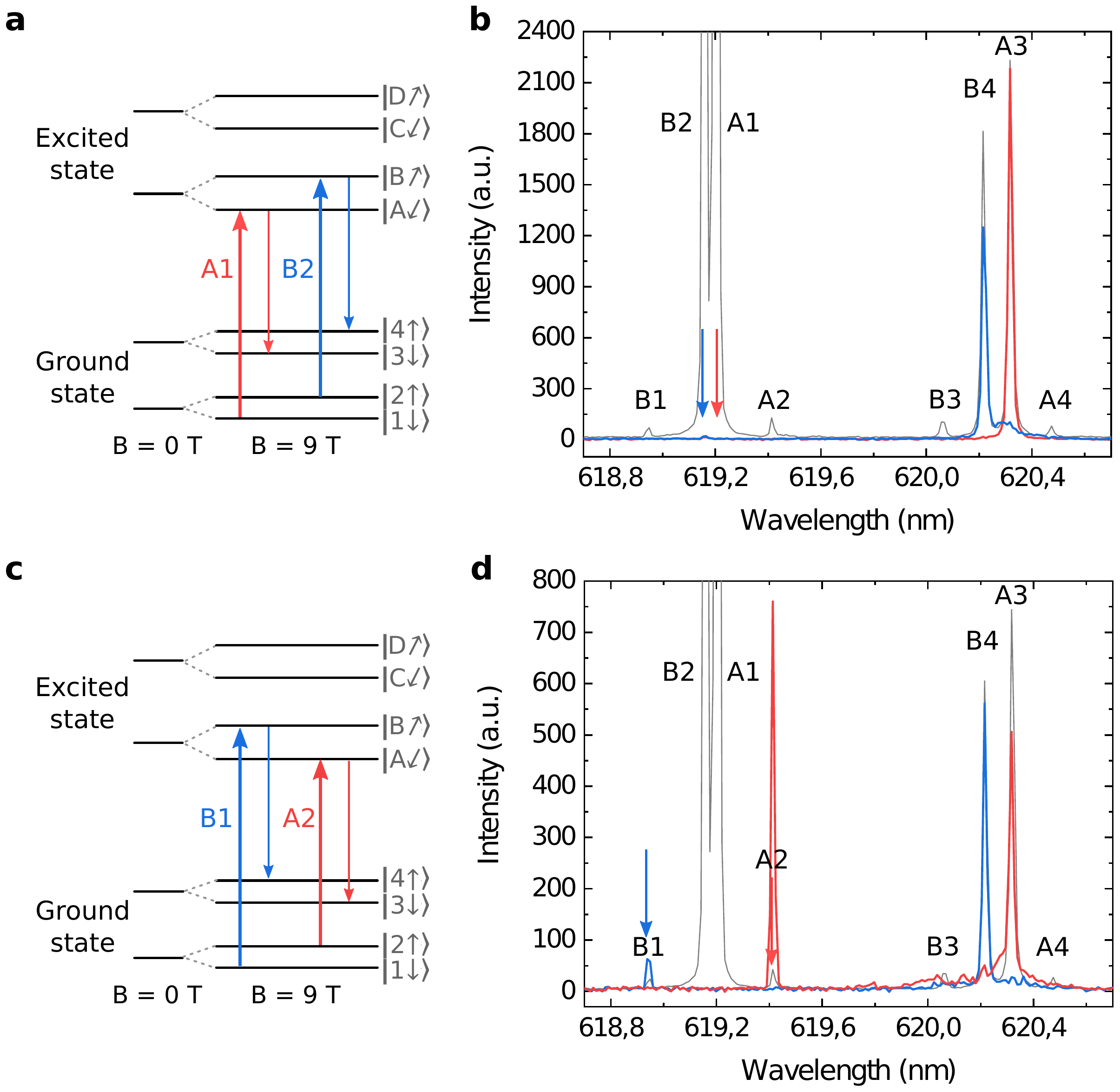}
    \caption{Spin-selective transitions. (a) SnV energy levels under external magnetic field depicting resonant excitation (upward arrows) and subsequent fluorescence (downward arrows) for transitions A1 (red) and B2 (blue) between levels of nearly overlapping spin states. (b) Experimental spectra at 9 T for non-resonant excitation (light gray curve) and resonant excitation of A1 (red curve) and B2 (blue curve). The excitation wavelengths are indicated by arrows of the corresponding colors. (c) Energy levels of the SnV center illustrating resonant excitation (upward arrows) and subsequent fluorescence (downward arrows) for transitions B1 (blue) and A2 (red) between levels of nearly opposite spin states. (d) Experimental spectra at 9 T for non-resonant excitation (light gray curve) and resonant excitation of B1 (blue curve) and A2 (red). Leakage from the laser is visible in both cases at the excitation wavelength.}
    \label{fig3}
\end{figure}

With the spin selection rules determined, we measure single-SnV spin properties through time-resolved fluorescence under an applied $0.13$ T magnetic field along the [001] axis~\cite{SnVSuppInfo}. Upon resonant excitation of transition A1, the observed phonon-sideband fluorescence decays with time as shown in Fig. 4a. This intensity decay indicates spin polarization as population is pumped from state $\ket{1\downarrow}$ (bright) to state $\ket{2\uparrow}$ (dark), reaching a spin initialization of 98$\%$ at a measured temperature of 4 K. Next, we measure the spin population decay time from the ratio of fluorescence at the leading edge of the initial polarizing pulse $I(0)$ to that of a second readout pulse delayed by a dark interval of length $\tau$ (Fig. 4b). The measured decay time $T_1 = 1.26\pm0.28$ ms is already an order of magnitude longer than for the SiV and GeV centers under similar magnetic field configuration and temperature~\cite{SnVSuppInfo, Rogers2014-qs, Pingault2017-gs, Becker2018-nv, Siyushev2017-lp}. The spin lifetime is expected to increase upon cooling since phonon-mediated orbital scattering is predicted to be the dominant decay pathway, as is the case for other group-IV centers. We confirm this hypothesis by measuring the change in $T_1$ with temperature, as shown in Figure 4c. The observed dependence follows the Bose-Einstein population of the phonon mode resonant with the orbital splitting of the SnV~\cite{SnVSuppInfo}, confirming that single phonon-mediated excitation to the upper ground state orbital branch is the dominant spin decay mechanism~\cite{Jahnke2015-pu,Pingault2017-gs} in this temperature range. Decreasing the temperature from 6 to 3.25 K the results in a spin lifetime reaching 10.4 ms, an exponential increase of over two orders of magnitude in good agreement with theoretical predictions.

To measure the phase coherence time, we implement optically detected magnetic resonance (ODMR), where we apply a microwave pulse between the optical initialization and readout pulses and scan its frequency across the transition between the two ground state spin states (levels 1 and 2). On resonance, the microwave pulse repopulates the initially depleted state $\ket{1\downarrow}$, thus leading to a peak in the spectrum. At 6 K, we observe a Lorentzian peak with full width at half-maximum $2.7\pm0.4$ MHz, indicating a modest spin coherence time $T_{2}^{*} = 59\pm8$ ns (Fig 4d). Upon further cooling to 2.9 K, leading to a concomitant reduction in phonon-induced dephasing, we observe a significant reduction in the ODMR linewidth down to $293\pm30$ kHz corresponding to a spin coherence time of $T_2^* = 540\pm40$ ns. This value is in good agreement with coherence being now limited by the $^{13}$C nuclear spin bath present in natural-abundance diamond\cite{Sukachev2017-sv,Sohn2018-bb}. We also reveal a double-peaked structure for this SnV center, likely indicating a strongly-coupled spin 1/2 system such as a nearby $^{13}$C\cite{Sohn2018-bb}. These results show a straightforward path for improvement. Our temperature-dependent spin measurements indicate that cooling SnV to 2 K is equivalent to cooling SiV to 100 mK in terms of reducing phonon-induced spin decoherence ~\cite{SnVSuppInfo, Sukachev2017-sv}. Coherent control of the SnV spin will allow the implementation of dynamical decoupling sequences that eliminate dephasing by the $^{13}$C nuclear spin bath, enabling spin coherence beyond milliseconds at these temperatures.

\begin{figure}
    \includegraphics[width =9cm]{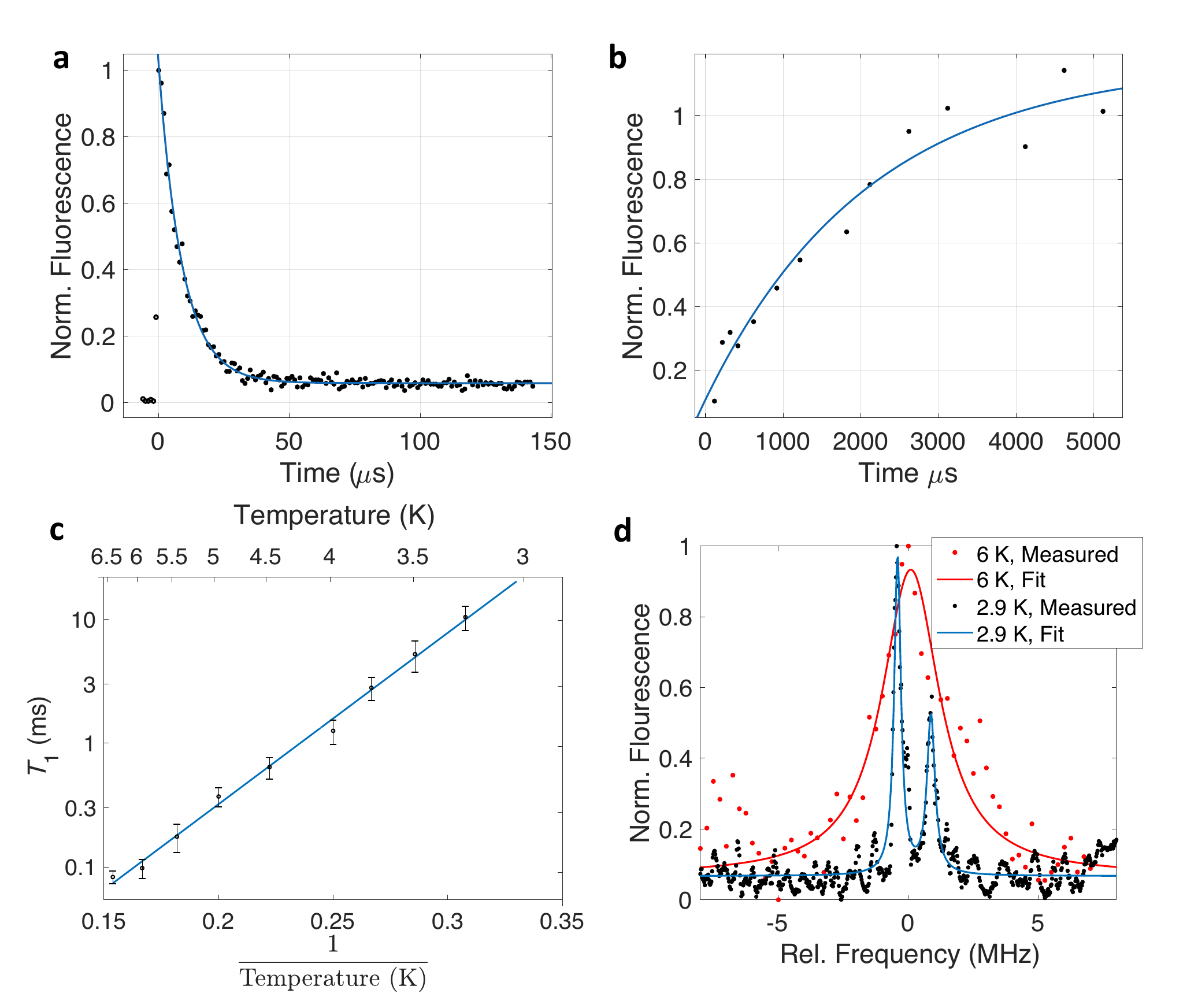}
    \caption{Spin properties of a single SnV. (a) Spin polarization under an applied magnetic field of $~0.13$ T along [001] at a measured temperature of 4 K. The phonon sideband fluorescence decays as a function of time under resonant pumping of the A1 transition as level 2 is populated. (b) Polarization decay. The fluorescence of the A1 transition recovers as a function of time between the excitation and readout pulses, indicating repopulation of the ground spin level 1. Blue curve: fit to an exponential with $T_1 = 1.26\pm0.28~$ms. (c) Spin decay time $T_1$ measured as a function of temperature (black dots). The blue solid curve is a exponential fit showing that $\log(T_1)$ varies linearly with 1/$T$, as expected for single phonon-induced spin decay~\cite{Jahnke2015-pu,Pingault2017-gs}. (d) Optically detected magnetic resonance spectra. The frequency of an applied microwave field is scanned across the transition between levels 1 and 2, leading to a repopulation of level 1 at resonance. The data (dots) are fitted with a Lorentzian functions (solid curve) with widths at half-maximum of $2.7\pm0.4$ MHz at 6 K (red), and $293\pm30$ and $356\pm60$ kHz at 2.9 K (blue).}
    \label{fig4}
\end{figure}

Our spin-resolved spectroscopy shows that the SnV center is a promising spin-photon interface combining transform-limited linewidths in nanostructures, a spin lifetime that far exceeds those of the SiV and GeV under the same conditions, and long spin dephasing times at accessible liquid-helium temperatures (obviating the requirement of dilution-refrigeration needed for SiV and GeV). These essential ingredients enable the development of quantum information processing and quantum photonic devices based on the SnV in a significantly more feasible manner. The addition of more elaborate nanophotonic structures for enhanced light-matter interaction~\cite{Schroder2016-bf} and coherent spin control for dynamical decoupling~\cite{Sukachev2017-sv} will  allow the development of SnV-based quantum network nodes. \\

\begin{acknowledgements}

M.T. acknowledges support by an appointment to the Intelligence Community Postdoctoral Research Fellowship Program at MIT, administered by Oak Ridge Institute for Science and Education through an interagency agreement between the U.S. Department of Energy and the Office of the Director of National Intelligence. B.P. acknowledges support from Wolfson College Cambridge through a research fellowship. N.H.W is supported in part by the Army Research Laboratory Center for Distributed Quantum Information (CDQI). K.C.C. acknowledges funding support by the National Science Foundation Graduate Research Fellowships Program (GRFP). E.B. was supported by a NASA Space Technology Research Fellowship. DL acknowledges support from the German Research Foundation (DFG) with a postdoctoral fellowship. D.E. and experiments were supported in part by the STC Center for Integrated Quantum Materials (CIQM), NSF Grant No. DMR-1231319. M.A. and experiments were supported in part by the University of Cambridge, the ERC Consolidator Grant PHOENICS (No. 617985), and the EPSRC Quantum Technology Hub NQIT (EP/M013243/1).

M.T and B.P contributed equally to this work.

\end{acknowledgements}

\bibliography{SnVbib.bib}

\end{document}